\documentclass[twocolumn,amssymb, nobibnotes, showpacs, superscriptaddress, aps, prd]{revtex4}
\usepackage{amsmath}
\usepackage{epsfig}
\begin{document}
\title{Strong optical self-focusing effect in coherent light scattering with condensates}
\author{Chengjie Zhu}
\affiliation{National Institute of Standards \& Technology, Gaithersburg, Maryland USA 20899}
\affiliation{East China Normal University, Shanghai, China 200062}
\author{L. Deng}
\affiliation{National Institute of Standards \& Technology, Gaithersburg, Maryland USA 20899}
\author{E.W. Hagley}
\affiliation{National Institute of Standards \& Technology, Gaithersburg, Maryland USA 20899}
\author{G.X. Huang}
\affiliation{East China Normal University, Shanghai, China 200062}

\date{\today}

\begin{abstract}
We present a theoretical investigation of optical self-focusing effects in light scattering with condensates. Using long ($>200\ \mu s$), red-detuned pulses we show numerically that a non-negligible self-focusing effect is present that causes rapid optical beam width reduction as the scattered field propagates through a medium with an inhomogeneous density distribution. The rapid growth of the scattered field intensity and significant local density feedback positively to further enhance the wave generation process and condensate compression, leading to highly efficient collective atomic recoil motion. 
\end{abstract}

\maketitle

{\it Introduction. $-$} Effects of a strongly-driven medium on the propagation of a near resonant light field
have been extensively studied in both linear and nonlinear optics. In linear optics, a
medium with a non uniform index of refraction, such as an optical fiber \cite{agarwal}, 
can lead to a lensing effect that causes the light field traversing the medium to be focused or defocused, depending on the detuning of the light with respect
to some general transitions of the medium.  In nonlinear optics \cite{shen}, however,
a significant local light field intensity can itself substantially alter the local optical index of refraction. This process, known as the Kerr effect, can result in laser beam self-focusing/defocusing, and even material break down and laser beam filamentation.  These effects have been widely observed both in gaseous phase and solid-state media at room temperature.  Theoretically, the general practice is to begin with the material equations without considering the center-of-mass motion (CM) of individual atoms or molecules participating in the wave generation and propagation process. This makes sense because in a room-temperature gaseous-phase medium the random thermal motion of the scatterers completely dwarfs any possible collective CM. In a solid-state medium, on the other hand, the scatterers are tightly bounded to their lattice sites, so again the CM motion is not important.  

\vskip 10pt

Self-focusing of an optical field in a medium is a non-linear process that arises from the local
change of the refractive index of the material induced by the intensity of an optical field.  In typical solid state material this often requires an intense electromagnetic field \cite{Cumberbatch1970,Mourou2006}. In room-temperature dilute gaseous phase media this effect is generally unimportant even with an intense parallel-beam light pulse of a relatively short pulse length.  This is, however, not the case with an ultra cold quantum gas where the extremely narrow optical transition line width between momentum states can lead to highly efficient generation of a light field within a very small propagation distance.  The spatial inhomogeneity of the density distribution of a trapped condensate, the extremely small medium cross section, and the confinement of a fast growing optical field result in an extraordinary optical self-focusing phenomenon that has never been seen before in a room temperature dilute gas.  We further note that in an ultra-cold quantum gas, such as a Bose condensate trapped in a magnetic trap, the collective CM recoil motion of atoms is of paramount importance.  This new feature leads to modified material equations and therefore phenomena that have not been examined previously.  

\vskip 10pt

In this work, we present a numerical study that investigates the optical self-focusing effect by considering both dynamic medium density evolution and the impact of local field growth due to an abnormally rapid local field cross section change. We first derive a (2+1)-D nonlinear Schr\"{o}dinger (NLS) equation from the Gross-Pitaevskii equation and the Maxwell equation describing the dynamic propagation effects due to an internally generated field in a Bose condensate by stimulated Raman scattering. We show by extensive numerical simulations that under long-pulse, red-detuned laser excitation significant coherent growth of the scattered field by a wave mixing process leads to a rapid reduction of the local field cross section and also results in a self-focusing effect that significantly alters the spatial inhomogeneity of a gaseous phase Bose condensate.  Before describing our work, we first point out that many early experimental \cite{Inouye1999A,Schneble2003,Schneble2004,Kuga2004,Inouye1999B,Kozuma1999} and theoretical \cite{Moore1999,Li2000,Piovella2001,Pu2003,Bonifacio2004,Fallani2005,Sarlo2005,Yu2004,
Uys2007,Benedek2004,Robb2005,Ketterle2001,Zobay2006,trifonov,sorenson,Deng2010A,Deng2010B,buchmann2010} studies have been devoted to light scattering in a Bose condensate.  These works, which mostly considered the linear regime of the scattering process, have contributed substantially to the understanding of the light scattering in condensates. 

\vskip 10pt

\begin{figure}
  \centering
  \includegraphics[width=7.5cm]{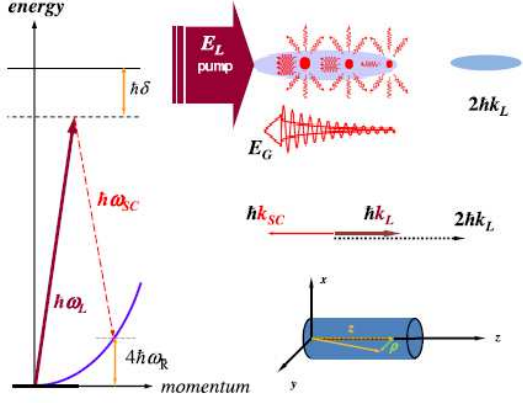}\\
  \caption{(Color online) Energy levels with laser couplings (left) and scattering geometry in a cylindrical coordinate system (lower-right). The red wavy arrow depicts the coherently scattered field with the largest gain. An atom absorbs a photon from the pump and then emits a photon via stimulated emission in the direction opposite to the pump, acquiring a net $2\hbar k_{\rm  L}$ momentum in the direction of the pump laser.}\label{fig:model}
\end{figure}

{\it Thoery. $-$} We start with a set of equations of motion describing the atomic mean field amplitudes and the propagation of the generated electric field inside the condensate. We consider a longitudinal pump scheme where a pump beam (field amplitude $E_L$) polarized in the $x-$direction propagates along the long axis of the condensate which is aligned with the $+z-$direction. In addition, a new field $E_{G}$, (see Fig. \ref{fig:model}) is generated inside the medium and it counter-propagates relative to the pump laser. More specifically, we assume that
%
\begin{eqnarray*}
\mathbf{E}_{L,G}^{(+)}&=&E_{L,G}^{(0)(+)}e^{i{\bf{k}}_{L,G}\cdot{\bf{r}}-i\omega_Lt}{\mathbf{e}}_x,\nonumber\\
\psi(\rho,z,t)&=&\sum_m\psi_m(\rho,t)e^{imKz-i\omega_mt},\nonumber
\end{eqnarray*}
where ${\bf{k}}_{L,G}\!\cdot\!{\bf{r}}\!=\!\pm k_{L,G}z$, $K\!=\!k_L+k_G$ and ${\mathbf{e}}_x$ is the polarization direction of the light fields. For what follows, we assume a uniform and constant pump $E_{L}^{(0)(+)}$ and a generated field of $E_{G}^{(0)(+)}=E_{G}^{(0)(+)}({\bf r},t)$. Without loss of generality, we also assume the condensate is cylindrically shaped and has a uniform density distribution along the long $z-$axis. However, the initial transverse density profile taken to be $n(\rho)=n_0(1-\rho^2/\rho_0^2)$ where $\rho^2=x^2+y^2$ ($r^2=\rho^2+z^2$) and $n_0$ is the peak density. Here, $\rho$ is the radial coordinate and $\rho_0$ is the initial transverse radius of the condensate (i.e., the short axis, see Fig. \ref{fig:model}). In the case of a true two-level system this longitudinal pump scheme is isomorphic to the transverse pumping scheme which yields two end-fire modes.

\vskip 10pt

With respect to Fig. 1, the equation of motion for the $n$-th order mean field atomic wave function is given by
\begin{eqnarray}\label{eq:GP}
\frac{\partial\psi_n}{\partial t}&=&i\frac{\hbar}{2M}\nabla_{\bot}^2\psi_n-iV_T\psi_n -ig_0\delta|\epsilon^{(+)}|^2\psi_n \nonumber\\
&-&ig\sum_{m_1,m_2}\psi_{m_1}\psi_{m_2}^\ast\psi_{n-m_1+m_2}S(n,m_1,m_2,t) \nonumber\\
&-&ig_0\delta\epsilon^{(-)}\psi_{n-1}e^{-i(\omega_{n-1}-\omega_n)t-i\Delta_L t} \nonumber\\
&-&ig_0\delta\epsilon^{(+)}\psi_{n+1}e^{-i(\omega_{n+1}-\omega_n)t+i\Delta_Lt},
\end{eqnarray}
where $S(n,m_1,m_2,t)=e^{i(\omega_n-\omega_{m_1}+\omega_{m_2}-\omega_{n-m_1+m_2})t}$, and $g=4\pi\hbar^2a/M$ with $a$ being the scattering length. In addition, $g_0=|D_{12}|^2|E_L|^2/(\hbar^2|\Delta|^2)$, where $\Delta=\delta+i\Gamma$ with $\delta$ and $\Gamma$ being the one-photon laser detuning to the upper electronic excited state and the spontaneous emission rate of the upper state, respectively. The normalized field is defined as $\epsilon^{(\pm)}=E_G^{(\pm)}(\rho,z,t)/E_L^{(\pm)}$, with $E_{L,G}^{(-)}=E_{L,G}^{(+)\ast}$. The trapping potential $V_T=M\Omega_T^2\rho^2/2$ with trapping frequency $\Omega_T$. $\hbar\omega_m=(m2\hbar k)^2/2M$ is the $m-$th order recoil energy with $k=k_L$ and $M$ being the pump laser wave vector and the mass of the atom, respectively. 

\vskip 10pt
In the slowly varying envelope approximation the Maxwell equation for the generated field is given by
\begin{eqnarray}\label{eq:Max_v1}
-i\frac{\partial\epsilon^{(+)}}{\partial z}&+&i\frac{1}{c}\frac{\partial\epsilon^{(+)}}{\partial t}+\frac{1}{2k_G}\nabla^2_\bot\epsilon^{(+)}=\frac{\kappa_0}{\Delta}|\psi_0|^2\epsilon^{(+)}\nonumber\\
&+&\frac{\kappa_0}{\Delta}\sum_n\psi_n\psi_{n+1}^\ast e^{i2(n+1)4\omega_Rt-i\Delta_Lt},
\end{eqnarray}
where the second term on the right is the polarization source term that drives the generation of the new field. In deriving Eq. (\ref{eq:Max_v1}) we have only kept the lowest scattering order, i.e. we neglect $n>1$ terms. Furthermore, we also neglect $n<0$ terms since it has already been shown that for long pulse excitation the bandwidth of the laser is sufficiently narrow that $n<0$ scattering orders do not occur.

\vskip 10pt

To investigate the scattered optical field self-focusing effect Eq. (\ref{eq:Max_v1}) must be solved simultaneously with the atomic response Eq. (\ref{eq:GP}) to third order in the generated field. We apply a perturbation expansion scheme 

\begin{eqnarray}
\psi_{0}=\psi_0^{(0)}+\lambda^2\psi_0^{(2)},\,
\psi_{1}=\lambda\psi_{1}^{(1)}+\lambda^3\psi_{1}^{(3)},
\epsilon^{(+)}=\lambda\epsilon^{(+)}.
\end{eqnarray}
These are well-known multi-scale pertubation schemes that have been widely used in soliton theories where small ground state population corrections must be included in the mathmetical theory \cite{soliton1}.  
Inserting Eq. (3) into Eq. (\ref{eq:GP}) we obtain

\begin{subequations}
\begin{eqnarray}
\frac{\partial\psi_0^{(0)}}{\partial t}&=&i\frac{\hbar}{2M}\nabla_{\bot}^2\psi_n^{(0)}-iV_T\psi_n^{(0)} -ig|\psi_{0}^{(0)}|^2\psi_{0}^{(0)}, \\
\frac{\partial\psi_0^{(2)}}{\partial t}&=&-\gamma_0\psi_0^{(2)}+i\frac{\hbar}{2M}\nabla_{\bot}^2\psi_0^{(2)}-iV_T\psi_0^{(2)}\nonumber\\ &-&ig_0\delta|\epsilon^{(+)}|^2\psi_0^{(0)}
-2ig|\psi_{+1}^{(1)}|^2\psi_{0}^{(0)}-ig|\psi_{0}^{(0)}|^2\psi_{0}^{(2)} \nonumber\\
&-&ig_0\delta\epsilon^{(+)}\psi_{+1}^{(1)}e^{-i\omega_{1}t+i\Delta_Lt},\\
\frac{\partial\psi_{+1}^{(1)}}{\partial t}&=&-\gamma_1\psi_{+1}^{(1)}+i\frac{\hbar}{2M}\nabla_{\bot}^2\psi_{+1}^{(1)}-iV_T\psi_{+1}^{(1)}\nonumber\\ 
&-&2ig|\psi_{0}^{(0)}|^2\psi_{+1}^{(1)}-ig_0\delta\epsilon^{(-)}\psi_{0}^{(0)}e^{i\omega_{1}t-i\Delta_Lt},\\
\frac{\partial\psi_{+1}^{(3)}}{\partial t}&=&-\gamma_1\psi_{+1}^{(3)}+i\frac{\hbar}{2M}\nabla_{\bot}^2\psi_{+1}^{(3)}-iV_T\psi_{+1}^{(3)}\nonumber\\ 
&-&2ig|\psi_{0}^{(0)}|^2\psi_{+1}^{(3)}-ig|\psi_{+1}^{(1)}|^2\psi_{+1}^{(1)}\nonumber\\
&-&ig\psi_{+1}^{(1)}\psi_0^{(0)}\,^*\psi_0^{(2)}-ig\psi_{+1}^{(1)}\psi_0^{(2)}\,^*\psi_{0}^{(0)}\nonumber\\
&-&ig_0\delta\epsilon^{(-)}\psi_{0}^{(2)}e^{i\omega_{1}t-i\Delta_Lt}.
\end{eqnarray}
\end{subequations}
It is clear that Eq. (4a), which is the zero-order equation for $n=0$ mean field wave fucntion $\psi_{0}^{(0)}$, is just the Gross-Pitaevskii equation in the absence of the external electric field \cite{note1a}.  In our calculation Eq. (4a) is solved numerically by directly numerical integration. 
\vskip 10pt
In the derivation of Eq. (4b-4d) we have introduced decay constants $\gamma_0$ and $\gamma_1$ to characterize the loss of coherence of the atomic center-of-motion states due to the interaction with the pump light field. In general, the total system population conservation in such a simple two-level model implies $\gamma_0^{(2)}\approx-\gamma_1^{(1)}$. This has been verified numerically. Finaly,
we neglected a constant Stark shift/dipole potential due to the pump field that can be removed by a trivial phase transformation without affecting the polarization source term in Eq. (\ref{eq:Max_v1}). 
\vskip 10pt
Enforcing the first-order Bragg scattering condition $\omega_1-\omega_0=4\omega_{\rm R}=\Delta_{\rm L}$, and consistently keeping all terms up to the third order in the generated field, the Maxwell equation for the generated field now becomes

\begin{eqnarray}\label{eq:Max_v2}
\frac{\partial\epsilon^{(+)}}{\partial z}&+&\frac{i}{2k_{\rm G}}\nabla^2_{\bot}\epsilon^{(+)}=i\frac{\kappa_0}{\Delta}\left(|\psi_0^{(0)}|^2\epsilon^{(+)}+\psi_0^{(0)}\psi_{+1}^{(1)\ast}\right)\nonumber\\
&&+i\frac{\kappa_0}{\Delta}\left[2{\rm Re}\left(\psi_0^{(0)}\psi_0^{(2)\ast}\right)+|\psi_{+1}^{(1)}|^2\right]\epsilon^{(+)}\nonumber\\
&&+i\frac{\kappa_0}{\Delta}\left(\psi_0^{(0)}\psi_{+1}^{(3)\ast}+\psi_0^{(2)}\psi_{+1}^{(1)\ast}\right).
\end{eqnarray}
Here, we have neglected the $(1/c)\left(\partial\epsilon/\partial t\right)$ term because the dominant propagation velocity comes from the polarization term \cite{Deng2010A}.
\vskip 10pt
Under the steady state approximation analytical expressions of $\psi_0$ and $\psi_{+1}$ can be obtained.
The first-order solution of the scattered component becomes
\begin{equation}\label{eq:psi_1(1)}
\psi_{+1}^{(1)}=-i\frac{\delta g_0\psi_0^{(0)}}{\gamma_1+ig|\psi_0^{(0)}|^2}\epsilon^{(-)}.
\end{equation}
\vskip 10pt
Using Eq. (\ref{eq:psi_1(1)}), we obtain
\begin{equation}\label{eq:psi_0(2)}
\psi_0^{(2)}=-i\delta g_0\psi_0^{(0)}\alpha|\epsilon^{(+)}|^2,
\end{equation}
where
\begin{equation}
\alpha\!=\!\frac{1}{\gamma_0+ib}\left[1\!+\!\frac{\delta g_0g|\psi_0^{(0)}|^2}{\gamma_1^2\!+\!g^2|\psi_0^{(0)}|^4}-i\frac{\delta g_0\gamma_1}{\gamma_1^2+g^2|\psi_0^{(0)}|^4}\right].
\end{equation}
Here, we have abrivated the second term on the right of Eq. (4b) as $\hbar b\equiv\hbar^2k_{\bot}^2/2M$. Physically, it is a small transverse kinetic energy of atoms in the zeroth-order condensate due to transverse light force compression. The third order correct $\psi_{+1}^{(3)}$ is given by
\begin{eqnarray}\label{eq:psi_1(3)}
&&\psi_{+1}^{(3)}=-\frac{\delta^2g_0^2\psi_0^{(0)}}{\gamma_1+ig|\psi_0^{(0)}|^2}|\epsilon^{(+)}|^2\epsilon^{(-)}\nonumber\\
&&\times\left\{\alpha+\frac{g|\psi_0^{(0)}|^2}{\gamma_1+ig|\psi_0^{(0)}|^2}\left[2{\rm Im}(\alpha)+\frac{\delta g_0}{\gamma_1^2+g^2|\psi_0^{(0)}|^4}\right]\right\}.\quad\;
\end{eqnarray}
\vskip 10pt
We now explain the rationale for the above outlined perturbation scheme where only the $\psi_{+1}$ order is considered.  Our calculations are aimed at providing a trackable derivation with an analytical solution that can capture the key physics.  It is for this reason that we limit our treatment to a pump light scattering rate of $R<80$ Hz.  In this regime only first-order scattering has been observed experimentally.  Although the $\psi_{+2}^{(2)}$ term, which is the leading contribution from the $\psi_{+2}$ term, is on the order of $|\epsilon^{(+)}|^2$ (similar to that of $\psi_{+1}^{(3)}$), we have neglected it in the above calculation because the residual multi-photon Doppler shift affects the scattering efficiency of a four-photon process (the $\psi_{+2}$ term) much more strongly than a two-photon process (the $\psi_{+1}$ term) for a given laser band width.  In fact, this energy mismatch due to a residual Doppler shift is the primary reason why even at higher pump powers the scattering orders higher than four are difficult to observe under long-pulse excitation \cite{ref29}.  We emphasize, however, that we have carried out directly numerical integration of Eqs. (4a)$-$(4c) and (5) without further approximation and the results agree well with the above steady state treatment.

\vskip 10pt
Substituting Eqs. (\ref{eq:psi_1(1)})-(\ref{eq:psi_1(3)}) into Eq. (\ref{eq:Max_v2}) we arrive at a third-order wave equation analogus to a (2+1)-D nonlinear Schr\"{o}dinger (NLS) equation where the 3rd-order nonlinear contribution can effectively balance the beam loss due to diffraction due to the condensate size effect, and result in an optical field self-focusing phenomenon. In our case, this (2+1)-D NLS equation can be written as
\begin{equation}\label{eq:NLS}
i\frac{\partial\epsilon^{(+)}}{\partial z}-\frac{1}{2k_{\rm G}}\nabla^2_{\bot}\epsilon^{(+)}+W|\epsilon^{(+)}|^2\epsilon^{(+)}=-\beta\epsilon^{(+)}.
\end{equation}
Here the linear absorption/gain term is given by
\begin{subequations}\label{eq:co_NLS}
\begin{eqnarray}
&&\beta\approx\frac{\kappa_0n}{\delta}\left(1-\frac{\delta g_0gn}{\gamma_1^2}+i\frac{\delta g_0}{\gamma_1}\right)\\
&&W\approx\frac{\kappa_0\delta g_0^2n}{\gamma_1^2}\left(3-\frac{5\delta g_0gn}{\gamma_1^2}\right)+2i\frac{\kappa_0\delta g_0^2n}{\gamma_1^3},\quad
\end{eqnarray}
\end{subequations}
where $n=|\psi_0^{(0)}|^2$ is the initial transverse density profile.  In deriving Eqs. (11a, 11b) we have assumed $b\ll|\gamma_0|$ for mathematics simplicity.  This assumption has been verified by direct numerical evaluation of the transverse kinetic energy $\hbar b$. 
\vskip 10pt
It has been shown previously \cite{agarwal,shen,Cumberbatch1970,Mourou2006} that the sign of Re$[W]$ given in Eq. (11b) leads to self-focusing/self-defocusing effects. 
Indeed, Eq. (11b) predicts that: (i) For red detunings (i.e. $\delta<0$) Re$[W]$ is always negative for typical experimental parameters (see below), and this will result in a reduction of the transverse dimension of the generated field. Thus, one expects to see reduced diffraction, and possibly a self-focusing effect. 
(ii) For blue detunings (i.e. $\delta>0$) Re$[W]$ is also negative for typical experimental parameters and therefore one also expects a self-focusing effect \cite{note} except the strength of the self-focusing effect is considerably weaker (that is, for typical experimental parameters we always find that $|\rm{Re}[W_{red}]|>|\rm{Re}[W_{blue}]|$).  Finally, for typical experimental parameters Im$[\beta]$ and Im$[W]$ are always positive for both red and blue detunings, indicating linear and nonlinear gains.
%
\begin{figure}
  \centering
  \includegraphics[width=7.5cm]{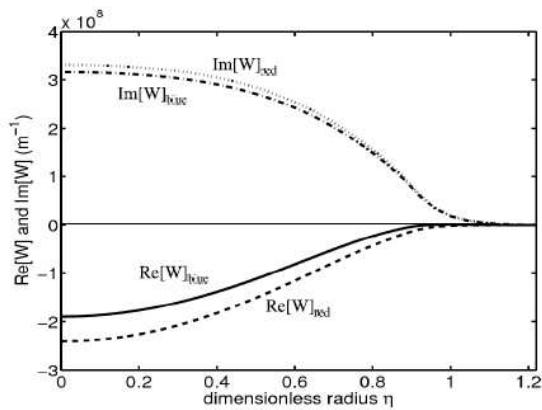}\\
  \caption{Third-order nonlinearity $W$ as function of $\eta=\rho/\rho_0$. Dashed line: Re$[W]_{\rm red}$, dotted line: Im$[W]_{\rm red}$ with red detunings $\delta/2\pi=-2$ GHz.  Solid line: Re$[W]_{\rm blue}$, dash-dotted line: Im$[W]_{\rm blue}$ with blue detunings $\delta/2\pi=+2$ GHz.}\label{fig:Wri}
\end{figure}
%
\vskip 10pt
{\it Numerical calculation. $-$} To verify the above analysis we performed full numerical simulations using Eqs. (10) and (11a,b).  Other parameters are s similiar to those reported in literature. Specifically, we consider a rubidium condensate with $2\times 10^6$ atoms, $L=200\ \mu$m, and $\rho_0=10\ \mu$m (peak density about $n_0=3.2\times10^{19}\ $m$^{-3}$).  $\Gamma/2\pi=6\ $MHz, $\gamma_1/2\pi=2$ kHz, $\gamma_0/2\pi=-2$ kHz, $\kappa_0=2.76\times10^{-6}\ {\rm m}^2{\rm s}^{-1}$,  $b=240$ Hz, $g/\hbar=4.85\times10^{-17}\ {\rm m}^3{\rm s}^{-1}$ corresponding to the scattering length $a_{\rm s}=100a_0$ (Bohr radius $a_0=5.29\times 10^{-9}$ cm), $\delta/2\pi=\pm 2$ GHz, $k_{\rm G}\approx 8\times 10^6$ m$^{-1}$. In accord with our approximations we chose $g_0=2.5\times10^{-5}$, which corresponds to $R\approx$ 60 Hz.  
In Fig. (2) we plot the values of Re$[W]$ and Im$[W]$ for these parameters.  It can be seen that indeed $|\rm{Re}[W]_{red}|>|\rm{Re}[W]_{blue}|$, and yet both contribute to a field self-focusing effect \cite{note}.
%
\begin{figure}
  \centering
  \includegraphics[width=7.5cm]{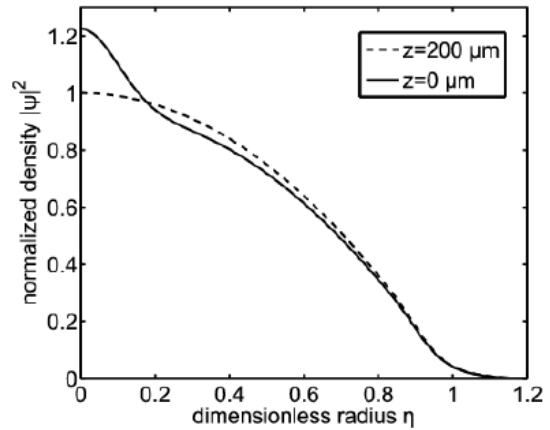}\\
  \caption{Macroscopic atomic mean field distribution as a function of dimensionless radius $\eta=\rho/\rho_0$ at $z=L$ (dashed curve) and at $z=0$ (solid curve). Note $z=L$ is the starting
position of $E_G$. At this point $E_G$ is negligible and the density distribution is just the original condensate distribution.  The field $E_G$ travels backward and it reaches its maximum value at $z=0$, causing the greatest atomic density change near the center of the condensate $\eta=0$. }\label{fig:p}
\end{figure}
%
\vskip 10pt
One important consequence of the light field self-focusing effect is its tendency to compress/decompress the spatial density distribution of the condensate.  This effect uniquely affects a gaseous phase medium where collective recoil motion is a prominent feature.  Indeed, such a density modification effect due to the light field intensity change is not important in a solid medium where the atoms are strongly bounded to their lattice sites.  Nor is this important for a normal gas where the collective CM recoil motion is completely negligible when compared to its intrinsic thermal motion.  In the case of red-detunings in a condensate, the self-focusing effect results in a rapid field intensity increase which further compresses the condensate.  This process further enhances the local field generation, resulting in positive feedback and a run away gain effect.  For blue-detunings, however, the atoms are expelled from the region of strong fields, resulting in a reduced density distribution which reduces the field generation efficiency.  In Fig. 3 we plot the atomic density distribution $|\psi(\rho,z)|^2$ as a function of the normalized radius $\eta$.  
We emphasize that the significant change in the local density distribution for red detunings shown in Fig. 3 further enhances the generation efficiency of the scattered light field, which further compresses the condensate.

\begin{figure}
  \centering
  \includegraphics[width=8.75cm]{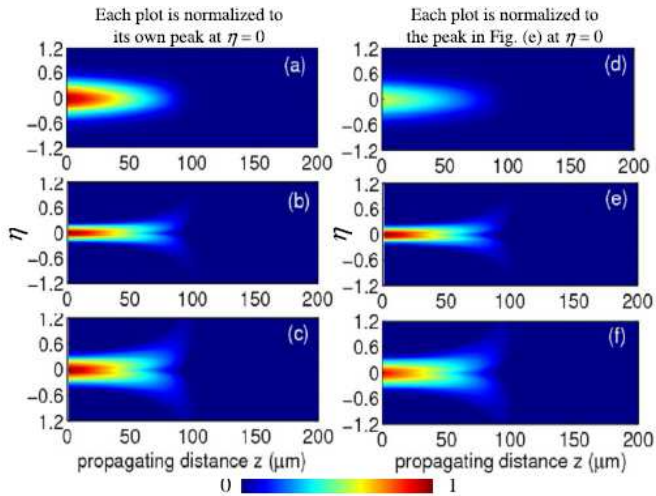}
  \caption{(Color online) Plot of $|\epsilon^{(+)}|^2$ as a function of the propagation distance $z$ and the dimensionless radius $\eta$. Left column: each plot is normalized to its own peak at $\eta=0$. Right column: all plots are normalized with respect to the peak of Fig. 4e at $\eta=0$ . Figs. 4a and 4d ($\delta/2\pi = -2$ GHz): The Kerr nonlinearity is neglected . Figs. 4b ($\delta/2\pi = -2$ GHz), 4c ($\delta/2\pi = +2$ GHz), 4e ($\delta/2\pi = - 2$ GHz), and 4f ($\delta/2\pi = +2$ GHz): The Kerr nonlinearity $W$ is included.}\label{fig:w}
\end{figure}
%
\vskip 10pt
This dramatic light field self-focusing effect is shown in Fig. 4 where the intensity profile of the generated light field is presented with, and without, the Kerr term for red and blue detunings.  Fig. 4a shows the field profile without the Kerr term ($\delta/2\pi = -2$ GHz).  Figures 4b and 4c show the field distributions with the nonlinear term included. Here, all three plots are normalized to unity to show the effective transverse field distribution (width).  Clearly, in the case of red-detuned pumps (Fig. 4b) the scattered field intensity has a cross section that is more than a factor of 2 smaller when compared to blue-detuned pumps (Fig. 4c), representing a factor of 4 \cite{fermion} intensity difference.  In Figs. 4d-4f we show the same numerical results but with all three plots normalized with respect to Fig. 4b.  This gives a sense of the relative strengths of the fields in Fig. 4a and 4c when compared to Fig. 4b. 

\vskip 10pt
{\it Conclusion. $-$} In conclusion, we have studied numerically the dynamic light field self-focusing effect in light scattering in a Bose condensate. By including the condensate transverse density profile we derived a 3-dimensional atomic CM Maxwell equation describing the generation and propagation of a new field, and a set of Gross-Pitaevskii equations for scattered atoms. Using a standard perturbation expansion, we recast the field equation into a (2+1)-D NLS equation which reveals the light field self-focusing phenomenon.  Numerical simulations revealed a significant reduction of the transverse profile of a red-detuned internally generated field as it propagates through the condensate. With red detunings the rapid increase in field intensity and the accompanying compression effect further feed back on themselves, leading to
a significant condensate density change and a highly efficient field generation and scattering process.  In the case of blue-detuned pumps, numerical calculations have shown that the field generation is considerably weaker. Our study, which provides the first theoretical evidence of nonlinear optical processes in light scattering in a condensate, has clearly shown that these higher-order processes play very important roles in light scattering in quantum gases.    

\vskip 10pt
Acknowledgments: Chengjie Zhu acknowledges supported by NSF-China under Grant Nos. 10874043 and 11174080, and by the Chinese Education Ministry Reward for Excellent Doctors in Academics under Grant
No. MXRZZ2010007.


\begin{references}
\bibitem{ref29} Higher-order scatterings occur in a sequential manner, implying that most of the key physics should be revealed by studying first-order scattering.
\end{references}

\end{document}